\documentclass[1p]{elsarticle}

\usepackage{lineno,hyperref}
\modulolinenumbers[5]
\journal{}
\usepackage{algorithm}
\usepackage{algpseudocode}
\usepackage{fix-cm}
\usepackage{mathtools}
\usepackage{amsmath,amsthm}
\usepackage{graphicx}
\usepackage{latexsym,enumerate,amssymb,amsbsy}
\usepackage{graphics,color}
\usepackage{verbatim}
\usepackage{url}
\usepackage{dblfloatfix}
\usepackage{mathptmx}
\usepackage{tikz-network}
\usepackage{booktabs}
\usepackage{xfrac}
\usepackage{amsfonts}
\usepackage{stmaryrd}
\usepackage{algorithm}
\usepackage{algpseudocode}
\usepackage{url}
\usepackage{colortbl}
\usepackage{adjustbox}
\usepackage{array,longtable}
\usepackage{caption}
\usepackage{subcaption}
\usepackage{arydshln}

\newtheorem{proposition}{Proposition}
\newtheorem{theorem}[proposition]{Theorem}

\newtheorem{example}{Example}

\newtheorem{definition}{Definition}

\newcommand\dsb[1]{\llbracket #1 \rrbracket}

\def\F{{\mathbb{F}}}

\def\Z{{\mathbb{Z}}}
\def\mdc{$G({\bf N}, S)$  }

\allowdisplaybreaks

\begin{document}
	
	\begin{frontmatter}
		\title{New Qutrit Codes from Pure and Bordered  Multidimensional Circulant Construction\tnoteref{t1}}
		\tnotetext[t1]{This material is based upon work supported by the National Science Foundation under Grant DMS-2243991.}

		\author[tamuc]{Padmapani Seneviratne\corref{cor1}}
		\ead{Padmapani.Seneviratne@tamuc.edu}	
		
		\author[TC]{Hannah Cuff}
		\ead{hannah.cuff@trincoll.edu}
		
		\author[CU]{Alexandra Koletsos}
		\ead{ak4749@columbia.edu}

		\author[SC]{Kerry Seekamp}
		\ead{kseekamp@smith.edu}

		\author[PU]{Adrian Thananopavarn}
		\ead{adrianpt@princeton.edu}

		\address[tamuc]{Department of Mathematics, Texas A\&M University-Commerce,\\
			2600 South Neal Street, Commerce, TX 75428.}
		
		\address[TC]{Department of Mathematics, Trinity College, 300 Summit Street Hartford, CT 06106.} 
		
		\address[CU]{Department of Mathematics, Columbia University, 2990 Broadway New York, NY 10027.}
		
		\address[SC]{Department of Mathematical Sciences, Smith College, 10 Elm Street Northampton, MA 01063.}
		
		\address[PU]{Department of Mathematics, Princeton University, Fine Hall, Washington Road Princeton, NJ 08544-1000. }

		\cortext[cor1]{Corresponding author.}

		\begin{abstract}
			We use multidimensional circulant approach to construct new qutrit stabilizer  $\dsb{\ell, 0, d}$ codes with parameters $(\ell, d) \in \{(51, 16), (52, 16), (54, 17), (55, 17), (57, 17)\}$ through symplectic self-dual additive codes over $\F_9$. In addition to these five new codes, we use  bordered construction to derive two more qutrit codes  with parameters  $(\ell, d) \in \{(53, 16), (56, 17)\}$ that improve upon previously best known parameters. 
		\end{abstract}
		
	\end{frontmatter}
	
	
	\section{Introduction}\label{Intro}
	In this work we study qutrit stabilizer codes from multidimensional circulant (MDC) graphs, and bordered MDC graphs. 
	MDC graphs are a generalization of the well known circulant graphs to multiple coordinates and was first introduced by Leighton in~\cite{Leighton}.
	Recently, in~\cite{Q2MDC}, Seneviratne et al., used MDC graphs to obtain new zero dimensional qubit stabilizer codes that improved previously best  known parameters. Additive self-dual codes over $\F_9$ have gotten much less attention compared to their counterparts over $\F_4$. 
	
	Let $\F_9 = \{0, 1, \omega, \ldots, \omega^7\}$,  where $\omega^2 = \omega + 1$ and $\omega^4 = -1$. An additive code $C$ ol length $n$ over $\F_9$ is a $\F_3$-linear subgroup of $\F_{9}^{n}$.
	The weight of a vector ${\bf v} \in \F_{9}^{n}$ is the number of nonzero entries of ${\bf v}$. The least nonzero weight of all codewords in $C$ is called the minimum distance of $C$. If $C$ is an additive code of length $n$ over $\F_9$ with minimum distance $d$ and size $3^k$, then $C$ is denoted by  $(n, 3^k, d)_9$. 
	The conjugation of an element $x \in \F_9$ is given by $\overline{x} = x^3$. The Hermitian trace inner product of two vectors ${\bf x, y} \in \F_{9}^{n}$ is defined by $\left \langle {\bf x, y} \right \rangle = \omega^{2}({\bf x}\cdot \overline{{\bf y}}- \overline{{\bf x}}\cdot {\bf y})$. The dual code $C^{\perp}$, with respect to the Hermitian trace inner product is defined by 
	$C^{\perp} = \{ {\bf u} \in \F_{9}^n \mid \left \langle {\bf c, u} \right \rangle = {\bf 0} \; \mbox{for all}\; {\bf c} \in C \}$.
	An additive  code is  self-dual if $C = C^{\perp}$. A self-dual code has parameters $(n, 3^n)$. 
	A code which is self-dual with respect to the Hermitian inner product $\langle \bf{x}, {\bf y} \rangle  = {\bf x}\cdot \overline{ {\bf} y}$ over $\F_9$, is also self-dual with respect to the trace Hermitian inner product~\cite{Danielsen2010}.

	\section{Qutrit codes from multidimensional circulant graphs}
	
	\begin{definition} \cite{Leighton}
		A MDC graph \mdc has the vertex set	$V(G) =\{(a_1,  \ldots, a_k): a_1 \in \Z_{n_1},  \ldots, a_k \in \Z_{n_k}\}$  and two vertices ${\bf a} = (a_1,, \cdots, a_k)$ and ${\bf b} = (b_1, , \cdots, b_k)$ are adjacent if and only if $(a_1 - b_1\ (mod\ n_1)), \cdots, a_k - b_k\ (mod\ n_k)) \in S$, where  $ {\bf N} = (n_1, n_2, \ldots, n_k)$, $S  \subset \Z_{n_1} \times Z_{n_2}\times \cdots \times \Z_{n_k}$, with $ S= -S$ and ${\bf 0} \notin S$.	
	\end{definition}
	
	\begin{example}
		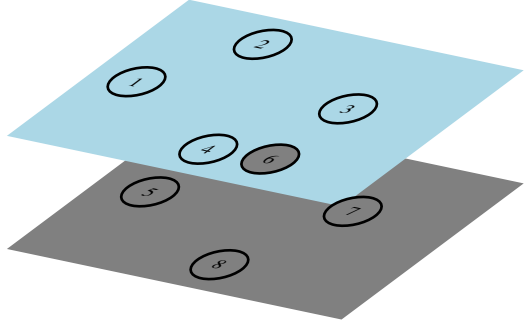
\begin{figure}[H] \label{fig:Q3}
			\centering
			\begin{tikzpicture}[multilayer=3d]
				\SetLayerDistance{-1.5}
				\Plane[x=-.7,y=-1.5,width=4.5,height=3,color=gray,layer=2,NoBorder]
				\Plane[x=-.7,y=-1.5,width=4.5,height=3,NoBorder]
				
				\Vertex[x=-0.1,y=-0.1, IdAsLabel,layer=1]{1}
				\Vertex[x=0.7,y=1,IdAsLabel,layer=1]{2}
				\Vertex[x=2.5,y=0.2,IdAsLabel,layer=1]{3}
				\Vertex[x=1.6,y=-1,IdAsLabel,layer=1]{4}
				\Vertex[IdAsLabel,color=gray,layer=2]{5}
				\Vertex[x=0.8,y=1,IdAsLabel,color=gray,layer=2]{6}
				\Vertex[x=2.4,y=0.4,IdAsLabel,color=gray,layer=2]{7}
				\Vertex[x=1.75,y=-1,IdAsLabel,color=gray,layer=2]{8}
				
				\Edge[color=blue](1)(2)
				\Edge[color=blue](2)(3)
				\Edge[color=blue](3)(4)
				\Edge[color=blue](4)(1)
				
				\Edge[style=dashed](5)(6)
				\Edge[style=dashed](6)(7)
				\Edge(7)(8)
				\Edge(8)(5)
				
				\Edge(1)(5)
				\Edge[style=dashed](2)(6)
				\Edge(4)(8)
				\Edge(3)(7)
				\Edge[style=dashed](6)(7)
			\end{tikzpicture}
			\caption{The $3$-cube graph as a MDC graph}
		\end{figure}
		Figure $1$ represents the hypercube graph $Q_3$ as a MDC graph with parameters $N=(2, 4)$ and $S = \{(0, 1), (0, 3), (1, 0)\})$. The upper layer contains the vertices in $V_0$ as $1, \ldots, 4$ and the lower layer presents the vertices in $V_1$ as $5, \ldots, 8$.
	\end{example}

	An adjacency matrix $\Gamma = \Gamma(G)$ of a MDC graph $G$ is a nested block circulant matrix as described in~\cite{Q2MDC}.
	
	\begin{theorem}\label{thm:nested}
		The adjacency matrix $\Gamma(G)$ of a MDC graph \mdc has nested block circulant form 
		\begin{equation*}
			\Gamma(G) = 
			\begin{pmatrix}
				A_{1,1} & A_{1,2} & \ldots & A_{1,{l-1}} & A_{1,l} \\ 
				A_{1,l} & A_{1,1} & \ldots & A_{1,{l-2}} & A_{1,{l-1}} \\ 
				\vdots& \vdots & \ddots & \vdots & \vdots \\ 
				A_{1,3} & A_{1,4} & \ldots & A_{1,1} & A_{1,2} \\ 
				A_{1,2} & A_{1,3} & \ldots & A_{1,l} & A_{1,1}
			\end{pmatrix} ,
		\end{equation*}   
		where $\mathbf{N}=(n_1,n_2,\ldots,n_k)$, with $n_1 \le n_2 \le \ldots \le n_k$ and each block $A_{1, j}$ is a $N_1 \times N_1$ submatrix of $\Gamma(G)$ for $1 \le j \le n_1 $, and 
		$N_1 = \frac{(n_1n_2 \cdot n_3 \cdots n_k)}{n_1}$, $N_2 = \frac{N_1}{n_2}$, $N_3 = \frac{N_2}{n_3}$, $\ldots$ , $N_k = \frac{N_{k-1}}{n_k}$. 
	\end{theorem}
	
	The following result stated in~\cite{Danielsen2010} ensures that the additive codes obtained from $3$-weighted graphs are self-dual with respect the Hermitian trace inner product. A graph 
	$G = (V, E, W)$ is  an $m$-weighted graph if  $W$ is a set of weights in $\F_m$ associated to each edge of $G$ with the vertex set $V$ and  $E\subseteq V\times V$ is the edge set of $G$.
	
	\begin{theorem}\cite{Danielsen2010}
		Let $\Gamma$ be the adjacency matrix of a $3$-weighted graph, $I$ the identity matrix, and a primitive element $\omega\in \F_9$. The additive code $C_{G}$ with the generator $\Gamma_{G} = \Gamma + \omega\cdot I$ is in standard from and is self-dual with respect to the Hermitian trace inner product.
	\end{theorem}
	
	When the length $n$ ranges from $1$ to $40$ we  ran an exhaustive search to generate every possible self-dual additive code over $\F_3$ that could be generated with MDC graphs. 
	We present a table that provide a comparative study between qutrit codes generated with MDC and optimal qutrit codes listed in Grassl's Table~\cite{Grassl:Q3table}.
	
	In the table~\ref{table:enum}, $n$ denotes the number of vertices (length of the additive code), $N$ signifies distinct sets, $d_{\textit{max}}^{t}(n)$ indicates the maximum minimum distance among self-dual additive codes generated by MDC graphs,  and $d_{max}(n,0)$ means the maximum known minimum distance among all qutrit codes with distance $n$ and dimension $0$.
	
	\begin{table} [H]
		\begin{center}
			\caption{Minimum distances of $0$-dimensional qutrit codes from MDC graphs.}
			\setlength{\tabcolsep}{3pt}
			\begin{tabular}{c c c c|c c c c}
				\toprule 
				$n$ & $N$ & $d_{\textit{max}}^{t}(n)$ & $d_{max}(n,0)$ & $n$ & $N$ & $d_{\textit{max}}^{t}(n)$ & $d_{max}(n,0)$  \\
				\midrule
				{1} & (1) & 1 & 1 & {21} & (3, 7) & 8 & 8-10\\[1.5pt] \noalign{\vskip  -0.055 cm} 
				
				{2} & (2) & 2 & 2 & {22} & (2, 11) & 9 & 9-10\\[1.5pt]
				
				{3} & (3) & 2 & 2 & {23} & (23) & 9 & 9-11 \\[1.5pt] \cdashline{1-8} \noalign{\vskip  0.04cm} 
				
				{4} & (4), (2,2) & 2 & {3} & {24} & (3, 8) & 9 & {10-12}\\
				{} & {} & {} & {} & {} & (12,2),(6,2,2) & 8 & {}\\ [1.5pt] \cdashline{1-8} \noalign{\vskip  0.04cm} 
				
				{5} & (5) & 3 & 3 & {25} & (25), (5,5) & 10, 9 & 10-12\\[1.5pt] 
				
				{6} & (2, 3) & 4 & 4 & {26} & (2, 13) & 10 & 10-12 \\[1.5pt] \cdashline{1-8} \noalign{\vskip  0.04 cm} 
				
				{7} & (7) & {4} & {4} & {27} & 27 & 10 & {10-12} \\
				{}&{}&{}&{}&{}&  (9,3),(3,3,3) & 9, 6 & {}\\ [1.5pt] \cdashline{1-8} \noalign{\vskip  0.04 cm} 
				
				{8} & (8),(4,2),(2,2,2) & 4 & 4 & {28} & (4, 7),(14,2) & 10  & 11-13 \\ [1.5pt] 
				
				{9} & (9),(3,3) & 4 & 5 &   {29} & (29) & 10 & 11-14 \\ [1.5pt]
				
				{10} & (2, 5) & 5 & 6 & {30} & (3, 10) & 10 & 12-14\\[1.5pt] 
				
				{11} & (11) & 5 & 5 & {31} & (31) & 11 & 11-14\\[1.5pt] \cdashline{1-8} \noalign{\vskip  0.04 cm} 
				
				{12} & (3, 4),(6,2) & 6, 4 & 6 & {32}& (32),(16,2) & 11 & {12-14}\\ 
				
				{ } & { }& { } & { } & { }  & (8,4) & 10 & {}\\
				{ } & { }& { } & { } & { } & (8,2,2), (4,2,2,2),(4,4,2) & 8 & {}\\
				
				{ } & { }& { } & { } & { } & (2,2,2,2,2) & 8 & {}\\[1.5pt] \cdashline{1-8} \noalign{\vskip  0.04cm} 
				
				{13} & (13) & 6 & 6  &  {33} & (33) & 12 & 12-15\\ [1.5pt] 
				
				{14} & (2, 7) & 6 & 6-7 & {34} & (2, 17) & 12 & 12-16\\ [1.5pt]
				
				{15} & (3, 5) & 6 & 6-8 & {35} & (5, 7) & 12 & 12-16\\ \cdashline{1-8} \noalign{\vskip  0.04cm} 
				
				{16} & (8,2) & 7 & {7-8} & {36} & [4, 9], [18,2],[12,3] & 12 & {13-16}\\ 
				{} & (16),(4,4) & 6 & {} & {} &  (6,6) & 10 & {}\\
				{} & (4,2,2),(2,2,2,2) & 4 & {} & {}& {}& {}& {}\\[1.5pt] \cdashline{1-8} \noalign{\vskip 0.04cm} 
				
				{17} & (17) & 7 & 7-8 & {37} & (37) & 13 & 13-17\\[1.5pt]
				
				{18} & (2, 9),(6,3) & 8, 6 & 8 & {38} & (2, 19) & 13 & 14-18\\[1.5pt] 
				
				{19} & (19) & 8 & 8-9 & {39} & (3, 13) & 13 & 13-18\\[1.5pt]
				
				{20} & (4, 5),(10,2) & 8 & 8-10 & {40} & (5, 8),(20,2),(10,2,2) & 13,12 & 14-18\\ [1.5pt]
				
				\bottomrule
			\end{tabular} 
			\label{table:enum}
		\end{center}
	\end{table}
	\subsection{New qutrit codes}
	
	We present five new qutrit codes $Q_{51}, Q_{52}, Q_{54}, Q_{55}$, and $Q_{57}$ obtained from MDC graphs by using a randomized search method. 
	Minimum distance of the codes $Q_{51}, Q_{52}, Q_{54}, Q_{55}$, and $Q_{57}$ were computed using the Magma computer algebra system~\cite{BOSMA1997}. 
	It took $8$ hours, $9$ hours, $12$ days, $11$ days, and $16$ days to verify the minimum distances of these codes.
	
	\begin{proposition}
		A new self-dual additive code $C_{51}$  with parameters $(51, 3^{51}, 16)_9$ and the resulting  $\dsb{51, 0, 16}_3$ qutrit code $Q_{51}$ is derived from the 
		the MDC graph $G_{51} := G((3, 17), S)$ with 
		\begin{small}

			\begin{equation*}
				\begin{split}
					&S:= \{ (1, 1), (1, 2), (0, 4), (1, 10), (1, 15), (0, 7), (1, 4), 
					(2, 13), (0, 9), (0, 1), (0, 16), (2, 2), (0, 8), \\ 
					&(1, 11), (0, 3), 
					(0, 11), (2, 15), (1, 16), (2, 11), (1, 12), (2, 1), (1, 9), (0, 5), 
					(0, 12), (0, 14),  (2, 8),\\
					& (2, 5), (0, 13), (2, 6), (0, 10), (1, 6), 
					(1, 13), (2, 4), (0, 6), (2, 16), (2, 7) \}
				\end{split}
			\end{equation*}
		\end{small}
		$Q_{51}$ improves the minimum distance of the $\dsb{51, 0, 15}_3$ code in the Grassl's qutrit code database~\cite{Grassl:Q3table} by $1$.
	\end{proposition}

	\begin{proposition}
		A new qutrit code $Q_{52}$ with parameters $\dsb{52, 0, 16}_3$ that improves the minimum distance of the best known code by $1$, is generated via an additive self-dual code $C_{52}$ obtained from the MDC graph $G_{52} := G((4, 13), S)$, where 
		\begin{small}
			\begin{equation*}
				\begin{split}
					&S:= \{ (3, 1), (3, 4), (0, 10), (1, 9), (3, 6), (3, 0), (1, 12), (0, 2), 
					(1, 3), (2, 5), (0, 11), (3, 9), (1, 0), (2, 8), (3, 8),\\
					&  (1, 7), (0, 3), (1, 
					4), (1, 5), (3, 10) \}
				\end{split}
			\end{equation*}
		\end{small}
		
	\end{proposition}
	
	\begin{proposition}
		The MDC graph $G_{54} := G((3, 18), S)$, where 
		\begin{small}
			\begin{equation*}
				\begin{split}
					&S:= \{ (1, 1), (1, 2), (1, 14), (1, 5), (1, 15), (0, 7), (2, 13), (2, 3), (0, 1), (0, 8), (0, 11), (2, 15), (2, 11), (1, 12),\\
					&  (1, 7), (2, 1), (0, 5), (2, 17), (1, 9), (0, 12), (1, 17), (2, 5), (0, 17),	(0, 13), (2, 9), (2, 6), (1, 3), (0, 10), (2, 12),\\
					&  (1, 6), (1, 13), (2, 4), (2, 16), (0, 6)
					\}
				\end{split}
			\end{equation*}
		\end{small}
		generates a new qutrit code $Q_{54}$ with parameters $\dsb{54, 0, 17}_3$ and improves upon the previous best known minimum distance given in~\cite{Grassl:Q3table} by $1$.
	\end{proposition}
	
	\begin{proposition}
		The MDC graph $G_{55} := G((5, 11), S)$, where 
		\begin{small}
			\begin{equation*}
				\begin{split}
					&S:= \{  (1, 1), (1, 2), (3, 3), (1, 10), (1, 5), (1, 4), (3, 6), (0, 9), (4, 3), (0, 1), (4, 4), (3, 2), (3, 10), (0, 2),\\
					& (1, 7), (4, 1), (2, 1), (1, 9), (3, 0), (4, 8), (4, 10), (2, 8), (2, 5), (4, 9), (2, 9), (1, 3), (0, 10), (3, 4), (2, 10),\\
					&  (4, 6), (2, 0), (4, 2), (3, 1), (1, 8), (4, 7), (2, 7)
					\}
				\end{split}
			\end{equation*}
		\end{small}
		generates the code $C_{55}$, with parameters $(55, 3^{55}, 17)$ and the corresponding qutrit code $Q_{55}$ has parameters $\dsb{55, 0, 17}_{3}$.
	\end{proposition}

	\begin{proposition}
		A new qutrit code $Q_{57}$ with parameters $\dsb{57, 0, 17}_3$ is generated via an additive self-dual code from the MDC graph $G_{57} := G((3, 19), S)$, where 
		\begin{small}
			\begin{equation*}
				\begin{split}
					&S:= \{ (0, 7), (0, 2), (2, 17), (2, 0), (0, 16), (1, 2), (0, 14), (2,	8), (1, 4), (0, 10), (1, 18), (0, 17), (0, 9), \\
					&(1, 11), (1, 15), (2, 4), (0, 12), (0, 5), (2, 5), (1, 8), (0, 6), (2, 13), (1, 0), (2, 15), (1, 6), (2, 18), (0, 13),\\
					& (2, 1), (0, 11), (0, 1), (2, 11), (1, 1), (1, 14), 	(0, 3), (1, 7), (2, 12), (0, 18), (0, 8)	
					\}
				\end{split}
			\end{equation*}
		\end{small}
	\end{proposition}

	\begin{table} [H]
		\begin{center}
			\caption{Properties of the Graphs}
			\setlength{\tabcolsep}{3pt}
			\begin{tabular}{c c c c c}
				\toprule 
				Graph & $d_{min}(G)$ & $v(G)$  & $\gamma(G)$ & $|Aut(G)|$ \\
				\midrule
				$G_{51}$ & $16$ & $36$ & $10$& $102$ \\
				$G_{52}$ & $16$ & $20$ & $5$ & $104$\\
				$G_{54}$ & $17$ & $34$	& $8$ & $108$ \\
				$G_{55}$ & $17$ & $36$ & $10$ & $110$\\
				$G_{57}$ & $17$ & $38$ & $9$ & $114$ \\
				$\overline{G_{52}}$ & $16$ & $27$ & $7$ & $104$\\
				$\overline{G_{55}}$ & $17$ & $28$ & $5$ & $110$\\
				\bottomrule
			\end{tabular} 
			\label{table:graph}
		\end{center}
	\end{table}

	Table~\ref{table:graph} gives the properties of the MDC graphs that generated new codes. All have diameter $2$ and girth $3$.
	Listed are the minimum distance $d_{min}(G)$ of the additive code $C :=C(G)$, the valency $v(G)$, the maximum clique size $\gamma(G)$, and the size $ |Aut(G)| $ of the automorphism group.

	\section{Bordered multidimensional construction}
	When studying self-dual codes, the bordered construction is a well-known technique employed by coding theorists to extend the length of a code. The bordered construction had been successfully applied in~\cite{GulKim, Saito2019} to classify additive self-dual codes over $\F_4$, and in~\cite{Seneviratne2}, the bordered construction was used to obtain two new qubit codes with parameters $\dsb{81, 0, 20}_2$ and $\dsb{94, 0, 22}_2$ from metacirculant graphs. In this section, we will study additive self-dual codes obtained from  bordered adjacency matrices of MDC graphs. 
	
	Let $A:=A(G)$ be the multidimensional circulant adjacency matrix of a MDC graph \mdc. A bordered MDC matrix 
	\begin{equation*}
		\overline{A} = 	
		\begin{bmatrix}
			0 & 1 & \cdots 1  \\
			1 & 	 &				\\
			\vdots &  A  &  	\\
			1 &		&
		\end{bmatrix}
	\end{equation*}
	is the adjacency matrix of the graph $\overline{G}$, which is the graph obtained by adding an extra vertex $v_{\infty}$ to the graph $G$ and connecting $v_{\infty}$ to all other existing vertices.

	We used a randomized search on bordered MDC graphs to obtain two new qutrit codes with parameters $\dsb{53, 0, 16}_3$ and $\dsb{56, 0, 17}_3$. The minimum distance of these codes were calculated using Magma~\cite{BOSMA1997}. 
	
	\begin{proposition}
		The bordered MDC graph $\overline{G}_{52}$ with $G_{52} = G((2, 26), S)$, where 
		\begin{small}
			\begin{equation*}
				\begin{split}
					&S =   \{ (1, 1), (1, 2), (1, 25), (0, 4), (1, 10), (1, 5), (1, 15), (0, 9), (0, 23), (1, 21), (1, 11), (0, 3), (1, 24),\\
					&  (0, 11), (1, 16), (0, 2), (0, 15), (0, 12), (0, 14), (0, 17), (0, 24), (0, 20), (1, 3), (1, 0), (1, 23), (0, 22), (0, 6) \}
				\end{split}
			\end{equation*}
		\end{small}
		generates a new $(53, 3^{53}, 16)_9$ additive self-code $C_{53}$. The corresponding $\dsb{53, 0, 16}_3$  code $Q_{53}$ improves the minimum distance of the best known code given in Table~\cite{Grassl:Q3table} by 1.
	\end{proposition}

	\begin{proposition}
		A new additive self-dual code $C_{56}$ with parameters $(56, 3^{56}, 17)$ and a corresponding new qutrit code $\dsb{56, 0, 17}_3$, $Q_{56}$ is generated by the bordered MDC graph 
		$\overline{G}_{56}$, with $G_{56} := G((5, 11), S)$, where 
		\begin{small}
			\begin{equation*}
				\begin{split}
					&S:= \{ (0, 4), (3, 3), (3, 7), (3, 6), (1, 4), (0, 7), (2, 3), (4, 3), (0, 8), (2, 2), (0, 3), (3, 5), (1, 9), (4, 8), (3, 9),(2, 8),\\
					& (2, 5), (4, 5), (3, 8), (2, 6), (1, 3), (3, 4), (1, 6), (2, 4), (4, 2), (1, 8), (4, 7), (2, 7) 	 
					\}.
				\end{split}
			\end{equation*}
		\end{small}
		$Q_{56}$ improves the minimum distance of the best known qutrit code given in Table~\cite{Grassl:Q3table} by $1$.
		
	\end{proposition}
	
	\section{Concluding Remarks}	
	We have shown that MDC graphs can be effectively used to construct optimal ternary quantum error-correcting codes. MDC construction enable us to generate longer length codes efficiently than the circulant construction. 
	
	\section*{Supplementary material}	
	
The certificates of minimum distance computations for qutrit codes of lengths $51, 52, 53, 54, 55, 56$ and $57$  are labeled {\tt MDT51output.txt, MDT52output.txt, BDT53output.txt, MDT54output.txt, MDT55output.txt, BDT56output.txt }  and {\tt MDT57output.txt}.

	\bibliographystyle{plain}
	
\end{document}